\documentclass[aps,prx,reprint,superscriptaddress,amsmath,amssymb,floatfix,nofootinbib,longbibliography]{revtex4-2}
\usepackage[T1]{fontenc}
\usepackage{orcidlink}
\usepackage{booktabs,microtype}
\usepackage{xcolor,graphicx,tikz,quantikz,pgfplots}
\usepgfplotslibrary{groupplots}
\usetikzlibrary{arrows.meta,positioning}
\pgfplotsset{compat=1.17}
\definecolor{CBBlue}{HTML}{4477AA}
\definecolor{CBOrange}{HTML}{CC6677}
\usepackage{hyperref}
\hypersetup{colorlinks=true,linkcolor=CBBlue,citecolor=CBBlue,urlcolor=CBBlue}
\newcommand{\Prb}{\mathbb{P}}

\pgfplotsset{journal/.style={axis lines=box,tick align=inside,xtick pos=both,ytick pos=both,font=\rmfamily\footnotesize,label style={font=\rmfamily\footnotesize},tick label style={font=\rmfamily\scriptsize},legend style={font=\rmfamily\scriptsize,draw=none},axis line style={line width=0.4pt},tick style={line width=0.4pt},scaled ticks=false,grid=none}}
\begin{document}
\title{Competing Fault Histories in Lattice-Surgery Decoding}
\author{Niharika Verma\,\orcidlink{0009-0008-0339-6588}}
\email{niharikachandraverma@gmail.com}
\affiliation{Department of Physics, Indian Institute of Technology, Patna}
\begin{abstract}
Different physical faults can produce the same detector record while
carrying opposite logical labels. We investigate the consequence of
this ambiguity for a lattice surgery parity measurement. In a distance-three
merge--split circuit, single-gate faults compete with reset/readout pairs
on three complete syndromes. Full-syndrome probabilities give a simultaneous
95\% lower confidence bound of $16.64$ parts per million on the cost of retaining the selected
single-fault decisions at one noise point. Changing those decisions is harmful at low noise and beneficial at higher noise. To test the physical explanation, we attenuate eight implicated channels
and compare the change in override effect with a matched non-target
suppression. The two interventions
change the decoder effect differently, resolving a contribution associated with those locations. Likelihood predictions fixed before sampling then select
useful decisions under mixed Pauli noise and after extending preparation
and recovery. The latter test gives a reduction of $13.20$ parts per million,
with simultaneous 95\% bounds $[5.55,20.84]$. The mechanism is limited in scope: the revised decisions give up certain single-fault corrections, and under the same witness-selection rule no candidate is identified in the fully expanded distance-five circuit. These results connect a finite-noise decoding trade-off to specific physical channels and test how far that explanation can be used to predict logical-parity performance.
\end{abstract}
\maketitle

\section{Introduction}
Lattice surgery implements logical parity measurements by merging and
splitting surface-code patches \cite{horsman,litinski,twist}. Repeated local measurements supply the record from which a decoder infers the logical
outcome. This record is physically incomplete: it identifies the syndrome of a fault
history, but generally not the history itself. Distinct
histories may therefore carry different logical labels even when every
measured detector agrees.

This ambiguity is familiar in quantum error correction
\cite{dennis,fowlerreview,gottesman2024surviving}. A decoder designed to correct a specified
set of faults must return the prescribed labels on their syndromes. A
decoder minimizing average logical error instead compares the total
probabilities of the competing logical classes. Maximum-likelihood methods
make that comparison explicit \cite{bravyi,heim,pryadko}; minimum-weight
matching offers an efficient approximation \cite{fowlerprocessing,sparse}.
Accounting for multiple explanations or circuit-level correlations can
improve decoding \cite{criger,fragile}. The two objectives need not select
the same decision at finite noise.

The question we take up is how this difference arises from the physical
circuit of a logical operation. A low-weight witness can establish that
opposite labels are possible, but cannot determine their total probability.
Conversely, a lower logical error rate does not identify which physical
channels made the improvement possible. Connecting these two descriptions
is particularly relevant during lattice surgery, where the stabilizer
measurements and their temporal boundaries change. Circuit-level surgery
analyses \cite{chamberland}, experimental logical operations \cite{erhard},
and studies of reset-dependent temporal faults \cite{geher} motivate
examining the actual extraction circuit rather than only its code geometry.

We study a concrete competition between a single gate fault and pairs of
reset or readout faults. Both explanations produce the same complete
syndrome, but their logical labels differ. At low noise, the single-fault
class is favored. At higher noise, the combined probability of the competing
class can make that same correction decision costly. Our aim is to determine
whether this competition explains an observed decision reversal and whether
it can predict what happens when the noise or temporal boundaries change.

The investigation follows that question. We identify the physical histories,
quantify their full-syndrome probability balance, and measure the effect of
changing only the associated decoder decisions. We then intervene on the
implicated channels and compare with a matched suppression elsewhere in the
circuit. Finally, predictions fixed before sampling are tested under mixed
Pauli noise and in a longer preparation--recovery sequence. The benefit is
small and comes with a loss of selected single-fault correction behavior.
Longer merged sequences also remove the candidates found by our selection
rule. These observations define the scope of the mechanism.

The contribution is this physical account and its controlled tests.
The distinction between correcting prescribed faults and minimizing average
logical error is established \cite{bravyi,heim,pryadko}. Approximate
logical-class likelihoods are also established tools \cite{truncated}. Here those ideas are used to
connect particular noise locations to a measured and prospectively tested
trade-off in a complete lattice-surgery circuit.

\section{Circuit and competing physical explanations}
\label{sec:circuit}
We consider a logical $XX$ parity measurement between distance-three
surface-code patches, joined by a one-column bridge. There are three rounds
before merging, three merged rounds, and three rounds after splitting.
Input preparation and final readout are included. The circuit, generated
with a fixed revision of FTPrimitiveBench \cite{ftbench}, contains 41 qubits,
152 detectors, and one tracked binary logical-parity label. Reference-entangled
checks of the ideal merge--split operation are described in
Appendix~\ref{app:validation}.

The physical model includes independent gate, reset, readout, and idle noise.
For the initial experiments, all probabilities scale with $\gamma$
according to Table~\ref{tab:noise}.
\begin{table}[t]
\caption{Circuit-level noise probabilities. All probabilities scale
with the dimensionless parameter $\gamma$.}
\label{tab:noise}
\begin{ruledtabular}
\begin{tabular}{lc}
Noise channel & Error probability \\
\hline
Two-qubit gate, $p_{2q}$                 & $10^{-3}\gamma$ \\
One-qubit gate, $p_{1q}$                 & $10^{-4}\gamma$ \\
Reset, $p_{\mathrm{reset}}$              & $0.012\gamma$ \\
Measurement, $p_{\mathrm{meas}}$         & $0.012\gamma$ \\
Idle, $p_{\mathrm{idle}}$                & $10^{-3}\gamma/6$ \\
Idle during measurement, $p_{\mathrm{idle,meas}}$
                                         & $10^{-3}\gamma/6$ \\
\end{tabular}
\end{ruledtabular}
\end{table}
Gate and idle faults are depolarizing; alternatives within one physical
location are mutually exclusive. Reset faults include initial preparation,
and readout faults flip reported measurement outcomes. The relative rates
make reset and readout faults appreciable in this study. They are specified
model parameters, not a calibration to a device.

Let $S$ denote the complete detector record and $L\in\{0,1\}$ the tracked
logical label. The three records of interest are
\begin{equation}
\mathcal A=\{D_{63}D_{77},\ D_{63}D_{80},\ D_{66}D_{80}\},
\label{eq:targets}
\end{equation}
with every unlisted detector bit zero. For each record, two identified
single-location Pauli faults carry $L=0$, whereas four reset/readout pairs
carry $L=1$. Retaining the complete record matters: the same two fired
bits inside a larger syndrome would define a different event and need not
have the probability balance studied here.

\begin{figure*}[t]
    \centering
    \includegraphics[width=\textwidth]{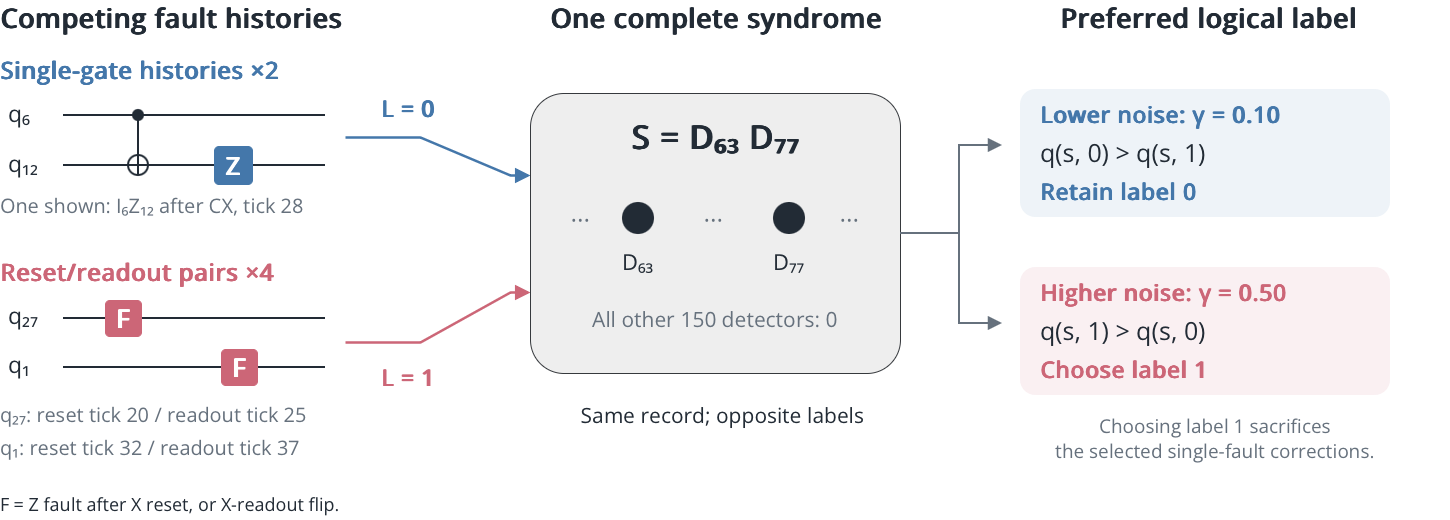} 
    \caption{Competing fault histories for the complete record $S=D_{63}D_{77}$,
with all other 150 detector bits zero. The upper branch shows one of two
single-gate histories carrying $L=0$; the lower branch represents four
reset/readout pairs carrying $L=1$. Each $F$ marker denotes either a
$Z$ fault after an $X$-basis reset or an $X$-readout flip: on $q_{27}$ at tick 20 or 25, and
on $q_1$ at tick 32 or 37, respectively. The class probabilities favor
label zero at $\gamma=0.10$ and label one at $\gamma=0.50$.
Changing the decision can reduce average parity error while sacrificing
correction of the identified single-gate faults. Circuit excerpts are
schematic; the full fault signatures were independently checked.}
    \label{fig:mechanism}
\end{figure*}

Figure~\ref{fig:mechanism} illustrates the first record. A fault $I_6Z_{12}$
after $\mathrm{CX}_{6\to12}$ at tick 28 gives only $D_{63}$ and $D_{77}$,
with label zero. The alternative $Z_{12}Z_{19}$ after
$\mathrm{CX}_{19\to12}$ at tick 29 has the same signature. A different
history chooses one reset/readout fault on $q_{27}$ at tick 20 or 25 and
one on $q_1$ at tick 32 or 37. Each of those four pairs produces the same
complete syndrome with label one. The six single-fault alternatives and
twelve competing pairs across $\mathcal A$ are verified by forced injections
and independent backward Pauli propagation.

We compare noise-matched minimum-weight perfect matching, $D_0$, with a
fixed modification $D_1$ that returns one on $\mathcal A$ and otherwise
agrees with $D_0$. In the original depolarizing sweep, matching returns zero
on these records and corrects all 2013 audited nonzero single-fault
signatures. The modification fails three. This loss is intentional: it
allows the cost of retaining those particular decisions to be measured.
Later experiments choose their labels from frozen likelihood predictions
rather than always returning one.

Our error metric is a wrong prediction of the tracked parity label. It is
not the error of the full logical quantum instrument. With that metric and
the two physical explanations specified, the remaining question is their
probability: an ambiguity alone does not tell us which decision is preferable.

\section{Correction constraints and finite-noise reversal}
\label{sec:cost}\label{sec:sweep}
\subsection{From fault histories to a correction cost}
Write $q(s,\ell)=\Prb(S=s,L=\ell)$. The logical-parity risk of a decoder is
\begin{equation}
 R(D)=\Prb[D(S)\ne L]=1-\sum_s q(s,D(s)).
 \label{eq:risk}
\end{equation}
The unrestricted Bayes decision chooses the larger class probability at
each syndrome; denote its risk by $R^*$. Any decoder required to retain label
zero on the selected records obeys
\begin{equation}
 R(D)-R^*\geq\sum_{s\in\mathcal A}[q(s,1)-q(s,0)]_+,
 \label{eq:lower}
\end{equation}
where $[x]_+=\max(x,0)$. This is a bound on the cost of the requirements,
independent of how the decoder acts elsewhere. Appendix~\ref{app:likelihood}
gives the general constrained-risk identity.

The comparison between the two implemented decoders is different. Their
risk difference is
\begin{equation}
 \Delta\equiv R(D_0)-R(D_1)
 =\sum_{s\in\mathcal A}[q(s,1)-q(s,0)].
 \label{eq:paired}
\end{equation}
Positive $\Delta$ favors the modification. Negative terms contribute to
Eq.~\eqref{eq:paired} but are omitted from Eq.~\eqref{eq:lower}; the
correction-cost bound is therefore not a forecast of the measured gain.
Both quantities include the rarity of the selected full records.

\begin{figure}[t]
\centering
\includegraphics[width=\columnwidth]{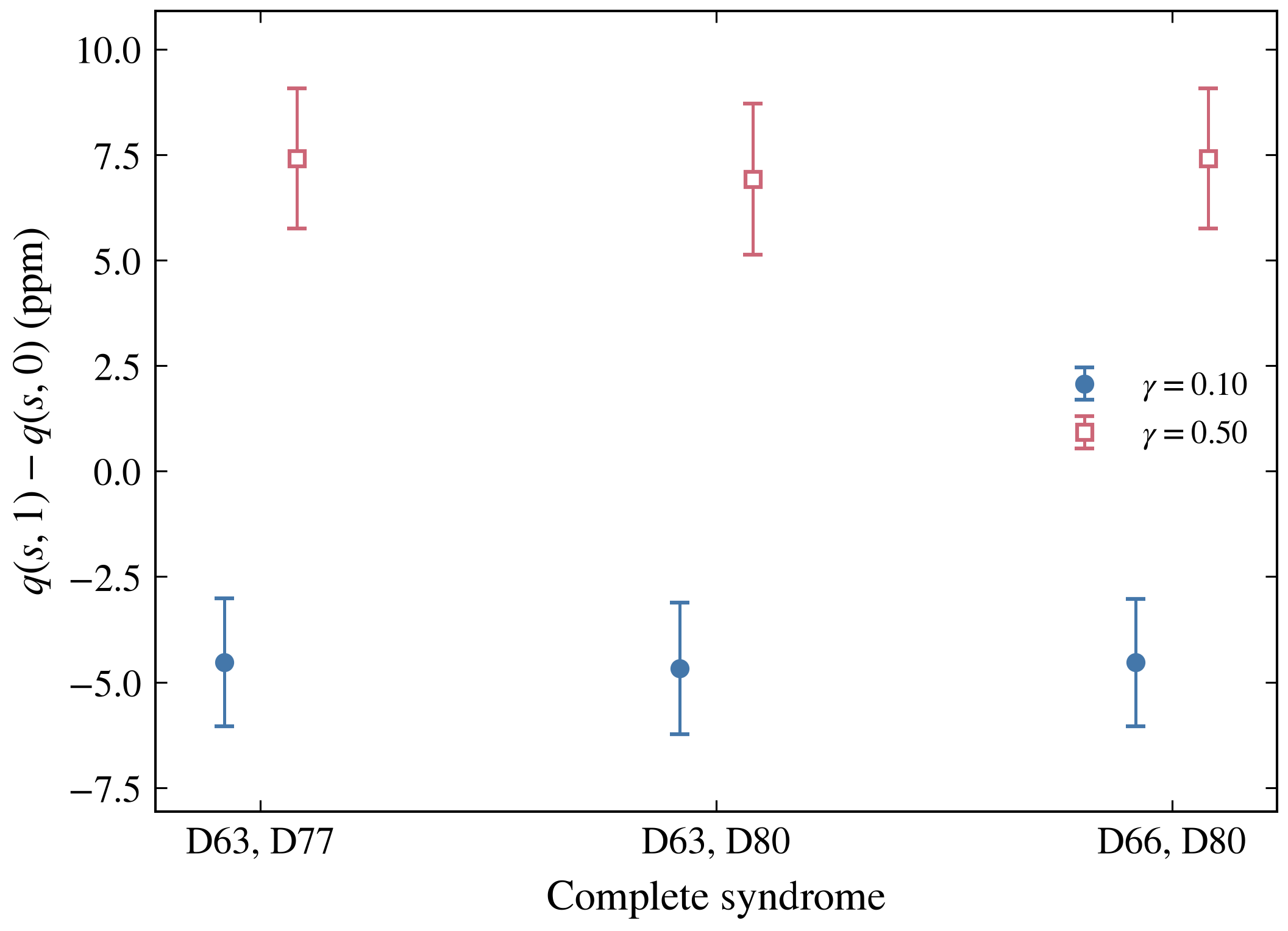}
\caption{Joint-probability differences for the three complete syndromes at two noise strengths. Intervals have simultaneous 95\% coverage in the expanded-support experiment, with ten million background samples per scale.}\label{fig:likelihood}\end{figure}

To evaluate the class probabilities beyond the identified low-weight
histories, we integrate 400 physical channels over a 14-detector support
and the logical bit, then sample the remaining 1498 channels without a
fault-order cutoff. Every sample contributes to the denominator, including
records with nonzero syndrome outside the support. Thus the calculation
retains full-syndrome probabilities. Ten million background samples at each
of $\gamma=0.10$ and $0.50$ give the simultaneous bounds in
Fig.~\ref{fig:likelihood}.

All three class differences are negative at the lower scale and positive
at the higher one. At $\gamma=0.50$, summing the positive lower endpoints gives
\begin{equation}
 R(D)-R^*\geq16.64\times10^{-6}
 \label{eq:numericalbound}
\end{equation}
with at least 95\% simultaneous statistical coverage, for decoders satisfying
the selected requirements. At $\gamma=0.10$, these records give a zero lower
bound. Neither statement characterizes all correction constraints or all
of matching's excess risk.

\subsection{A fixed decision change reverses its effect}
We next evaluate $D_0$ and $D_1$ on identical fresh circuit samples. There are
twenty million shots at each of seventeen noise scales from $0.10$ to $0.50$
in steps of $0.025$, with matching weights recomputed for each physical model.
The initial six-scale experiment was followed by eleven additional scales,
chosen after inspecting the initial results and fixed before the new sampling.
The override itself is unchanged. We report the combined grid with
retrospective simultaneous intervals shared with the intervention conditions
(Appendix~\ref{app:statistics}).
A paired comparison counts repairs of matching failures and failures
introduced by the modification; their difference estimates $\Delta$.
Simultaneous intervals are constructed from these discordant outcomes
rather than from independent estimates of the two total error rates
(Appendix~\ref{app:statistics}).

\begin{figure}[t]
\centering
\includegraphics[width=\columnwidth]{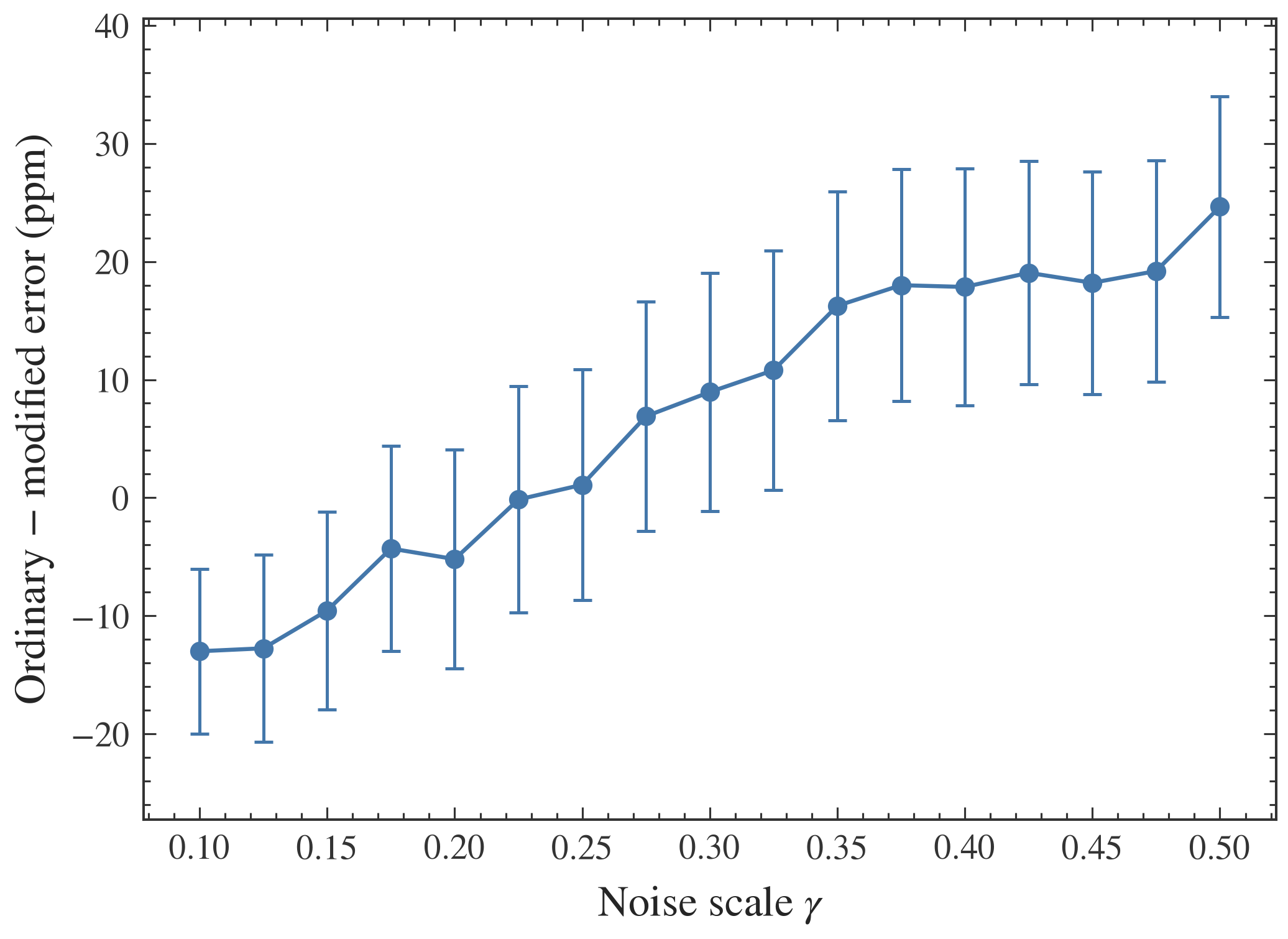}
\caption{Paired override effect at seventeen noise scales, with twenty million shots per point. Positive values favor the modification. Intervals use the retrospective simultaneous 95\% family of 27 displayed conditions and ten binomial marginals per condition. Six scales reuse the initial sweep; eleven are later additions. Lines connect sampled conditions, not fitted values. The points from $\gamma=0.175$ through $0.30$ are unresolved.}
\label{fig:sweep}\end{figure}

At $\gamma=0.10$, the modification increases parity error:
$\Delta=-13.00$ ppm, with simultaneous 95\% bounds $[-19.97,-6.03]$ ppm.
At $\gamma=0.50$, the sign reverses: $\Delta=+24.65$ ppm, with bounds
$[15.29,34.01]$ ppm. The displayed family resolves harm through $0.15$
and benefit from $0.325$ onward; the sampled points from $0.175$ to $0.30$
remain unresolved [Fig.~\ref{fig:sweep}]. The resolved signs and continuity
of the finite-circuit class probabilities support at least one zero between
$0.15$ and $0.325$. They do not establish a unique crossing. The broader
family makes the interval at $0.30$ include zero, although its estimate
is positive.

The gain is modest. At the highest scale, the ordinary and modified failure
rates are $0.560315\%$ and $0.557850\%$, a relative reduction of about
$0.440\%$. The $16.64$ ppm constraint bound and the $24.65$ ppm observed gain
come from separate calculations and answer different questions. Together,
they show that retaining the selected single-fault decisions can carry a
finite-noise cost, and that changing those decisions affects an implemented
decoder in the expected direction.

The reversal is consistent with the competing histories, but does not identify
its physical source. Many higher-order histories enter the same complete
records. We therefore test whether changing the implicated reset/readout
locations alters the decision balance differently from removing comparable
noise elsewhere.

\section{Testing the mechanism by targeted intervention}
\label{sec:intervention}
The identified pairs involve eight physical channels: four post-$R_X$
$Z$-error channels and four $M_X$ readout-flip channels. Each has probability
$0.006$ at $\gamma=0.50$. We multiply only those probabilities by a retained
fraction $\alpha$, keeping the ideal circuit and all other physical channels
fixed. The initial five fractions $\{0,0.25,0.50,0.75,1\}$ were extended
by $\{0.125,0.375,0.625,0.875\}$ after the initial results. The endpoints
correspond to complete suppression and the original noise model.

Removing eight channels could change performance simply by reducing noise.
To distinguish that possibility from an effect associated with their
locations, we also suppress eight non-target channels selected before
sampling. They match the targets in channel type, probability, tick,
extraction round, and surgery phase. Spatial separation is minimized with
deterministic tie-breaking. Detector support, graph degree, and boundary
role are not matched; these imperfections are retained in the interpretation.
Appendix~\ref{app:controls} lists the exact choices.

\begin{figure}[t]
\centering
\includegraphics[width=\columnwidth]{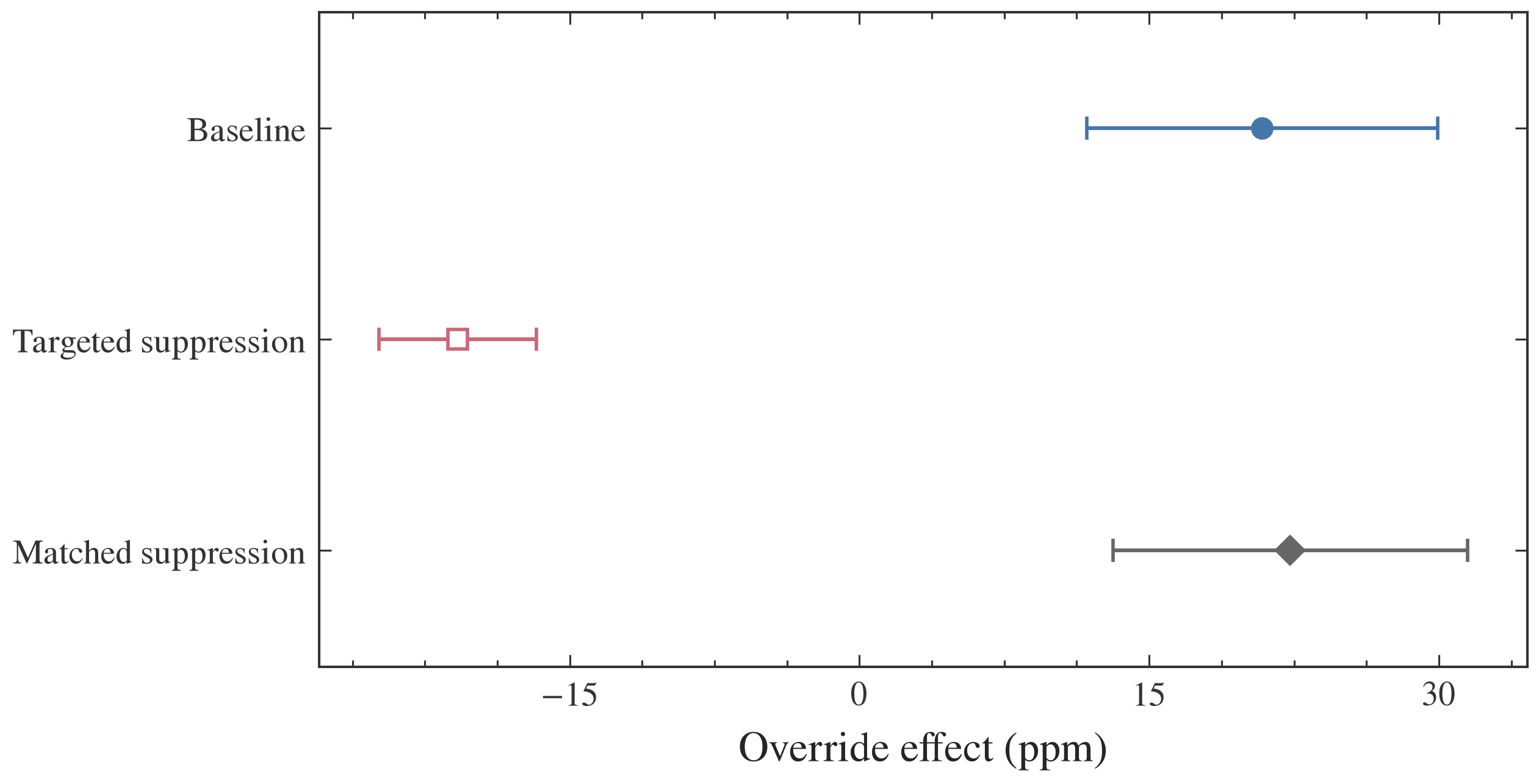}
\caption{Override effect for baseline, targeted suppression, and matched suppression. Each condition uses twenty million samples, independent of the noise sweep. The estimates reuse the original control experiment; their displayed intervals use the broader retrospective $27\times10$ family shared with the dense sweeps.}
\label{fig:matched-2}\end{figure}

Each of the nine attenuation conditions and the matched control receives
twenty million shots. The four added fractions use fresh samples; the
original six conditions retain their original samples. Matching
weights are recalibrated in each condition, while the same three decisions
are modified. The independent baseline gives $\Delta=+20.85$ ppm. Targeted
suppression changes the effect to $-20.80$ ppm, whereas matched suppression
gives $+22.30$ ppm [Fig.~\ref{fig:matched-2}]. The principal contrast is
\begin{equation}
 \Delta_{\mathrm{target}}-\Delta_{\mathrm{matched}}
 =-43.10\ \mathrm{ppm},
 \label{eq:contrast}
\end{equation}
with simultaneous 95\% bounds $[-56.36,-29.86]$ ppm. The two suppression
conditions remove the same sum of marginal channel probabilities, $0.048$;
this sum is not the probability that any removed fault occurs.

The attenuation experiment connects the targeted endpoints
(Appendix~\ref{app:controls}). Its point estimates change from negative to
positive as the selected noise is restored. The expanded display resolves
negative effects through $\alpha=0.50$ and positive effects from $0.875$;
$0.625$ and $0.75$ remain unresolved. Matched suppression differs from baseline by
$+1.45$ ppm, with bounds $[-16.80,19.70]$ ppm. That interval does not establish
equivalence to the baseline, and the attenuation points do not prove
monotonicity or a unique crossing.

The targeted-minus-matched contrast supports a location-specific explanation
within this model. It does not isolate exactly-two-fault trajectories:
suppressing a channel also removes every higher-order history containing it.
Nor does one matched set control all structural differences between
locations. Because matching is reweighted under each intervention, the
contrast describes the stated decoding procedure, rather than a derivative
at fixed edge weights. These qualifications leave a useful physical result:
removing the proposed channels changes the decision balance differently
from removing an equal amount of the same types of noise elsewhere.

This intervention tests the explanation at the original operating point.
We now ask whether it can do more than account for that result: can it select
useful decisions before outcomes are sampled under altered noise or circuit
boundaries?

\section{Prospective predictions and circuit dependence}
\label{sec:prospective}
\subsection{Predicting the decision under mixed noise}
For each target record, we calculate approximate class probabilities by
integrating all local fault orders within the 14-detector witness-channel
support and retaining at most two background faults. Background histories
contribute only when their outside detector syndrome cancels. Writing $a$
for the complete local law and $b_{\leq2}$ for this unnormalized background
law gives
\begin{equation}
 \widetilde q(s,\ell)=\sum_v a\bigl((s,\ell)\oplus v\bigr)b_{\leq2}(v).
 \label{eq:prospective}
\end{equation}
The predicted label maximizes $\widetilde q$ on each target; all other
decisions remain those of matching. This differs from the all-order
calculation underlying Eq.~\eqref{eq:numericalbound}. The discarded mass is
recorded: at $\gamma=0.50$, the unconditioned omitted mass is $0.402$
for the original circuit and $0.516$ for the boundary extension below.
These bounds are too broad to certify the signs. We therefore test
the approximation through predictions saved before fresh confirmation,
without treating agreement as an error guarantee. Details and numerical
crosschecks are given in Appendix~\ref{app:forecasts}.

The new conditions vary conditional Pauli composition while holding total
gate and idle fault probabilities fixed. The $X,Y,Z$ weights are
$(0.8,0.1,0.1)$, $(0.6,0.2,0.2)$, $(0.45,0.275,0.275)$, their $X/Z$
reflections, and the isotropic point. Reset and readout channels retain their
basis-specific form. Three noise scales, $0.10,0.25,0.50$, give eighteen new
mixed-bias conditions and three isotropic controls. Matching is calibrated
to each model. These experiments concern stochastic Pauli noise, not coherent
rotations.

\begin{figure}[t]
\centering
\includegraphics[width=\columnwidth]{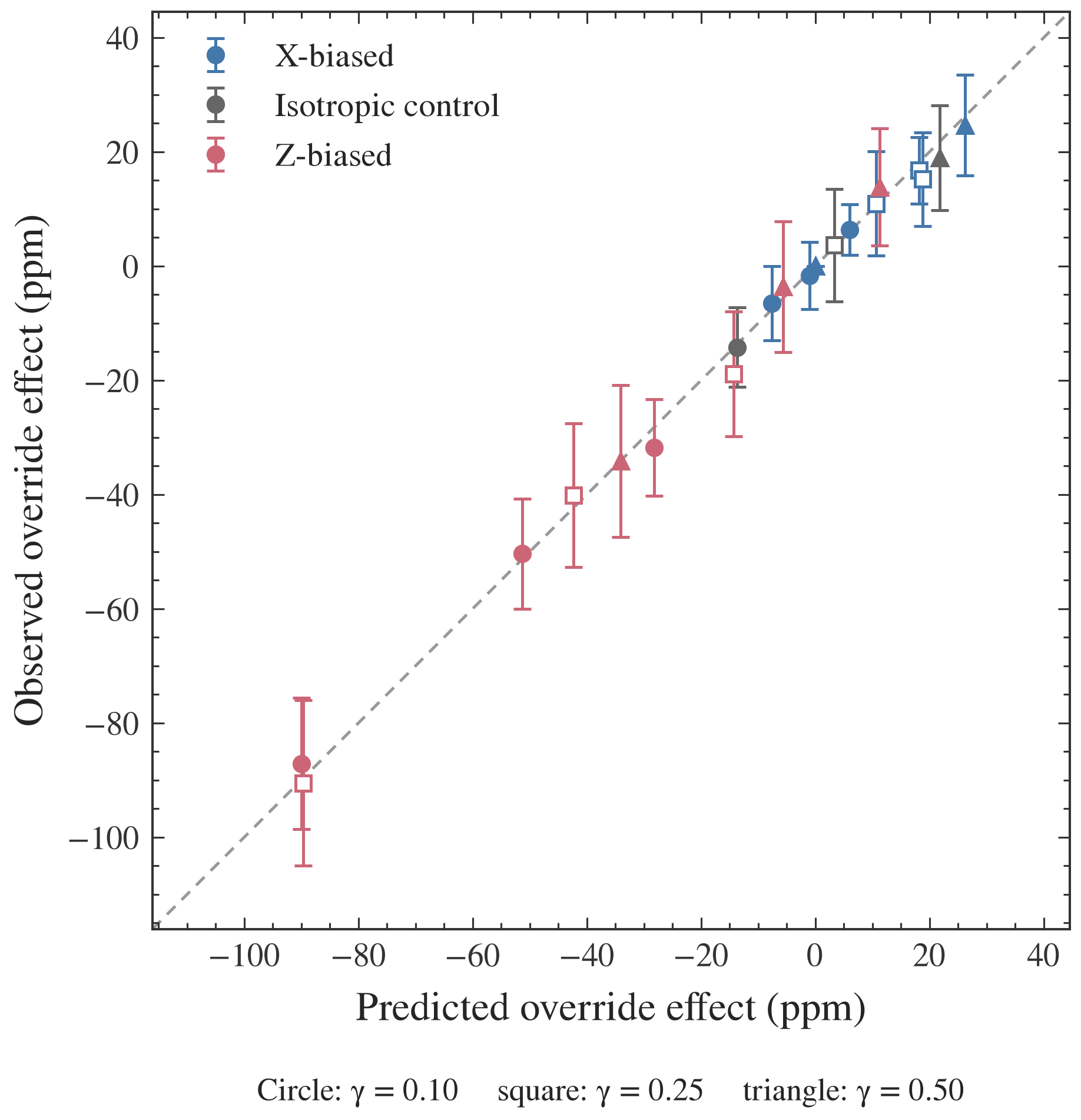}
\caption{Prediction versus confirmation for the fixed label-one rule in eighteen mixed-bias conditions and three isotropic controls. Color distinguishes the noise composition; circles, squares, and triangles denote $\gamma=0.10,0.25,0.50$. Each point uses twenty million fresh shots. Intervals retain the original simultaneous $21\times14$ confirmation family. The dashed line is equality, not a fit. Two exact policy identities coincide at the origin. These points evaluate the fixed rule. Within the eighteen mixed-bias conditions, the prediction-selected policy instead gives six resolved benefits and twelve identities.}
\label{fig:mixedprediction}\end{figure}

All predictions were fixed before twenty million fresh shots per condition,
for 420 million shots in total. Figure~\ref{fig:mixedprediction} compares the
predicted and observed effects of the fixed label-one rule for all twenty-one
conditions, including the isotropic controls and cases where the rule is harmful. The prediction-selected policy
improves six of the eighteen new conditions and is exactly identical to
matching in the other twelve. In contrast, always choosing label one is
harmful in eight conditions. Across the eighteen mixed-bias points, the
fixed rule reaches a harm of $90.50$ ppm, compared with a largest benefit
of $24.65$ ppm. The asymmetry matters: much of the prediction-selected
policy's value is in declining to apply a harmful override. The twelve
identities are abstentions, not twelve additional improvements. Every aggregate forecast lies within its
simultaneous confirmation interval. The result supports predictive utility
on this finite grid; it does not certify the truncation or establish
performance under arbitrary bias.

Correlated matching provides a second decoder reference
\cite{fowlercorrelated,pymatching}. Replaying the same 420 million records
and adding the same frozen decisions again gives six resolved improvements
and twelve identities in the mixed-bias subset. Both matching methods assign
the same labels to the three targets in all twenty-one conditions, so these
incremental effects must agree on shared samples. The comparison is useful
as a baseline check, but supplies no independent replication. The decoders
differ elsewhere and neither is uniformly better overall
(Appendix~\ref{app:correlated}).

\subsection{Extending preparation and recovery}
Changing the noise leaves the circuit fixed. To test a change in the circuit,
we increase preparation and recovery from three to four rounds each, while
retaining distance-three patches and three merged rounds. The extended
circuit has 184 detectors instead of 152. The unchanged witness-selection
rule finds nine candidates and selects three with a translated local support.
The extra rounds add background fault opportunities without changing the
local pattern of competing histories.

\begin{figure}[t]
\centering
\includegraphics[width=\columnwidth]{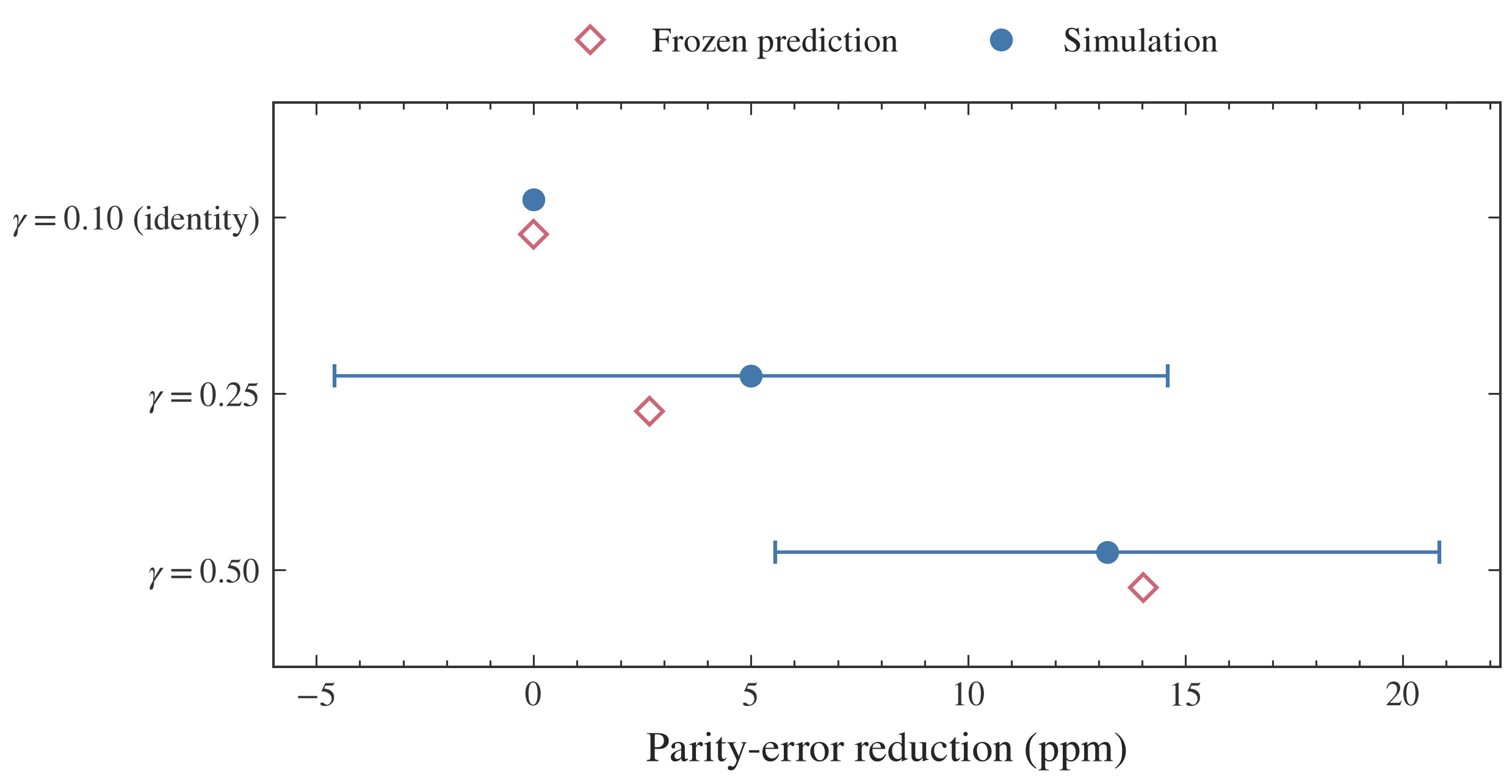}
\caption{Prospective test in the circuit with four preparation, three merge,
and four recovery rounds. Vertical offsets separate the two markers at
each noise scale; they have no physical meaning. Each scale uses ten million fresh shots. Error bars give simultaneous
95\% intervals for the three-condition confirmation family.}
\label{fig:transfer}\end{figure}

Target lists and predictions at the same three noise scales are fixed before
thirty million fresh shots, ten million per scale. At $\gamma=0.10$, the
selected policy leaves matching unchanged. At $0.25$, the predicted effect
is $2.67$ ppm and the observed effect is $5.00$ ppm with bounds
$[-4.59,14.59]$ ppm. The intermediate sign is unresolved. At $0.50$, the
prediction is $14.02$ ppm and the observed reduction is $13.20$ ppm,
with simultaneous 95\% bounds $[5.55,20.84]$ ppm
[Fig.~\ref{fig:transfer}]. These intervals share one confidence family for
the three new conditions.

At the highest scale, ordinary and modified matching make 57246 and 57114
errors, a relative reduction of about $0.231\%$. Ordinary matching corrects
all 2461 nonzero single-fault signatures audited in this circuit. The changed
rule fails three at both scales where its labels differ. Thus the extended
circuit retains both the small average benefit and its correction cost.
It tests a moved temporal boundary within the same circuit family, rather
than a new code or a different local mechanism.

The result does not extend automatically to longer merges. The selection
rule finds no candidate with four merged rounds, or when patch size and all
three stage depths are increased to five. Physical propagation checks and
ideal-operation validation support the latter structural result
(Appendices~\ref{app:transfer} and~\ref{app:fully-enlarged});
no noisy modified-decoder comparison is claimed
there. These absences concern the specified single-fault versus reset/readout-pair
construction. They leave open other syndromes and higher-order competitions.

\section{Discussion and conclusion}
The physical ambiguity is simple: a single gate fault and reset/readout
pairs can leave identical detector records with opposite logical labels.
What matters for decoding is the probability of each whole logical class.
The full-syndrome calculation shows that their balance changes across the
tested noise range, and the paired decoder experiment measures the
consequence of changing the corresponding decisions. A requirement that
corrects a particular fault can therefore increase average parity error
at finite noise, even when its physical justification at low noise remains
valid.

The intervention is what ties this observation to the proposed locations.
Suppressing the implicated reset/readout channels reverses the benefit of
the fixed rule, while suppressing a prospectively matched set leaves a
positive effect. The resolved difference between those interventions goes
beyond a decoder ranking. It associates the decision conflict with a
physical feature of the extraction circuit, subject to the unmatched
structural properties and the removal of higher-order histories described
in Sec.~\ref{sec:intervention}.

The prospective tests ask whether that account is useful beyond the setting
that suggested it. On the finite mixed-noise grid, the likelihood calculation
selects beneficial changes or retains the baseline. A fresh extension of
preparation and recovery also gives a small resolved benefit at the highest
noise level. These tests support the explanation within a defined range.
They do not turn the approximate probability calculation into a certificate:
its omitted-background bound is loose, and the longer-merge tests expose a
structural limit of the selected mechanism.

There is no conflict with asymptotic fault correction. In a fixed circuit
with consistent protection through $t$ faults, a protected history of weight
$w\leq t$ contributes at order $p^w$, whereas an opposite-label explanation
requires at least $t+1$ faults. The required label is then favored for
sufficiently small $p$, when all nonidentity outcome probabilities scale
with $p$. Our reversal concerns finite noise. It neither demonstrates a
threshold improvement nor supplies a conclusion uniform in distance.

The practical limitations are substantial. The measured gains are small,
selected single-fault corrections are sacrificed, and the independent Pauli
model is not calibrated to hardware. Leakage, coherent noise, and temporally
correlated environments are outside the benchmark. The metric also tracks
one logical-parity label rather than the full noisy quantum instrument.
A Hadamard-conjugated circuit is a symmetry check, not evidence that these
limitations disappear for another operation.

The study nevertheless gives a concrete way to examine a decoding trade-off:
identify the physical histories behind an ambiguous complete record,
quantify their probability balance, intervene on the proposed channels, and
test predictions before sampling new outcomes. Here that chain of evidence
supports a specific mechanism and also shows where its present construction
stops. That is the scope of the result.

\section*{Data and code availability}
The simulation and analysis code, circuit files, and selected numerical
data are available at
\url{https://github.com/nihoniumqc/reset-placement-study}.
The repository includes condition parameters and aggregate paired counts
for the reported decoder comparisons, including the seventeen-scale noise
sweep and nine-fraction attenuation sweep, together with the calculation
of their displayed confidence intervals. Retrospective attenuation
predictions are also included. Large likelihood-estimator checkpoints
are not part of this selected release.

\section*{Acknowledgment}
The simulation workflows used in this study were developed with the assistance of OpenAI Codex for initial coding, code review, and debugging.

\bibliography{bib_sources/biblio}
\clearpage
\twocolumngrid
\appendix
\section{Full-syndrome likelihood calculation}
\label{app:likelihood}
The integrated detector support is
\begin{equation}
\begin{split}
U=\{60,63,64,66,69,70,74,77,79,80,83,89,93,94\}.
\end{split}
\end{equation}
The local state retains these 14 detector bits and the logical bit, giving
32,768 states. A whole physical channel is integrated only when every one
of its detector outcomes is supported in $U$. For categorical outcome masks
$v_e$ with probabilities $p_e$, the local distribution is updated by
\begin{equation}
\lambda'(v)=\Big(1-\sum_e p_e\Big)\lambda(v)
+\sum_e p_e\lambda(v\oplus v_e).
\end{equation}
This preserves exclusivity between outcomes of a depolarizing location.
Of 1,898 channels, 400 are integrated and 1,498 remain in the background.
No fault-order cutoff is applied to either component.

Write $B_U$ for the background mask on $U$ together with the logical bit,
and $B_{\bar U}$ for its remaining detector bits. With
$\pi_b=\Prb(B_{\bar U}=0,B_U=b)$, independence gives
\begin{equation}
q(s,\ell)=\sum_b\pi_b\lambda((s|_U,\ell)\oplus b).
\end{equation}
Every sampled background contributes to the denominator, including those
with $B_{\bar U}\ne0$. The calculation therefore retains the full-syndrome
probability rather than a probability conditioned on passing an outside
syndrome check.

For each binomial marginal $\pi_b$, we construct two-sided Clopper--Pearson
bounds $[l_b,u_b]$, allocating failure probability
$0.05/(2\cdot32768)$ over both noise conditions and all categories. A union
bound gives simultaneous coverage without assuming independence between
category counts. For
$c_b=\lambda((s|_U,1)\oplus b)-\lambda((s|_U,0)\oplus b)$,
the lower bound on the probability difference is
\begin{equation}
\sum_{c_b\geq0}c_bl_b+\sum_{c_b<0}c_bu_b,
\end{equation}
with the upper bound obtained by interchanging interval endpoints.
Positive lower endpoints can be summed across the distinct syndromes in
Eq.~\eqref{eq:lower}. The unrounded result at $\gamma=0.5$ is 16.639558 ppm.
The local integration is exact for the finite categorical model in exact
arithmetic; its floating-point implementation is crosschecked rather than
formally enclosed by interval arithmetic.

\subsection{Correction requirements and joint risk}
For consistent required decisions $g(s)$ on $\mathcal S_F$, let $R_F^*$ be
the minimum risk subject to those decisions. Minimizing independently at
each syndrome gives the standard identity
\begin{equation}
R_F^*-R^*=\sum_{s\in\mathcal S_F}
\left[\max_\ell q(s,\ell)-q(s,g(s))\right].
\label{eq:cost}
\end{equation}
Setting $g(s)=0$ on $\mathcal A$ yields Eq.~\eqref{eq:lower}. Additional
requirements can only increase the constrained minimum risk.

\subsection{Conditional preference and absolute cost}
The role of rare syndromes is clearer if the class difference is written as
\begin{equation}
\begin{split}
q(s,1)-q(s,0)=\Prb(S=s)\big[&\Prb(L=1\mid S=s)\\
&-\Prb(L=0\mid S=s)\big].
\end{split}
\label{eq:conditional}
\end{equation}
A strong conditional preference can still have a small effect on the total
logical error if the syndrome is rare. The bound and decoder comparison use
the joint probabilities, so this rarity is included automatically. Neither
quantity is a conditional failure rate restricted to the selected events.

For a single selected syndrome, write
$\delta_s=q(s,1)-q(s,0)$. If $\delta_s>0$, retaining label zero contributes
$\delta_s$ to the excess risk relative to the Bayes decision. If
$\delta_s<0$, retaining label zero is already optimal for that syndrome,
and changing it introduces a cost of $-\delta_s$. With several selected
syndromes, these signed effects can cancel in the actual decoder comparison.
They cannot cancel in the constraint lower bound, whose terms are
$[\delta_s]_+$. This explains why the two reported quantities answer
different questions even though they are computed from the same class masses.

The lower bound is also local in its requirements. It applies to every
decoder required to keep the specified labels, regardless of how well that
decoder performs elsewhere. It does not measure all the excess risk of
ordinary matching. Other syndromes may contribute additional loss, and
nothing in the calculation establishes that the modified decoder attains
the unrestricted Bayes risk.

\section{Paired decoder outcomes}
\label{app:decomposition}
\begin{figure}[t]
\centering
\includegraphics[width=\columnwidth]{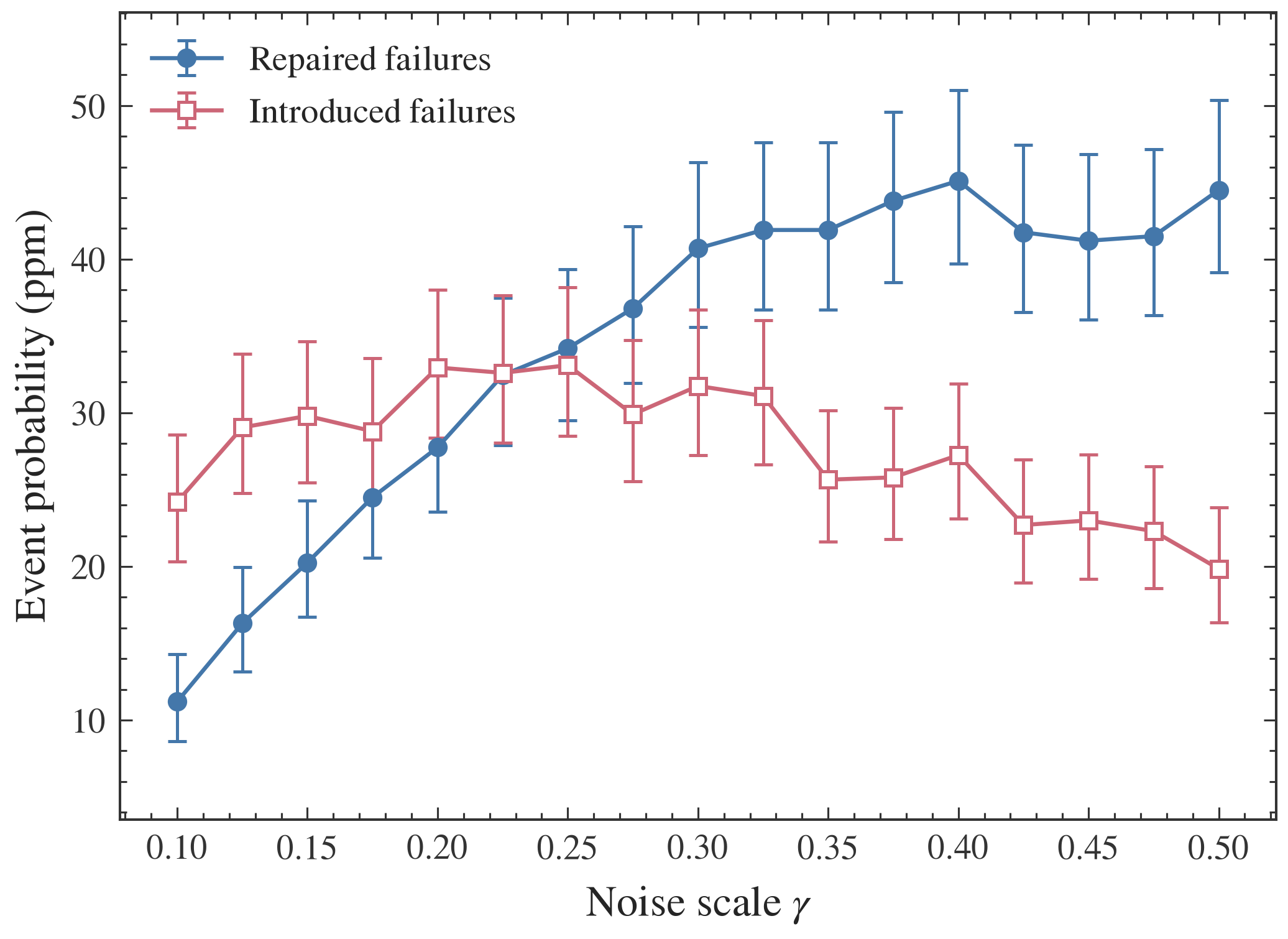}
\caption{Repair and introduced-failure probabilities at the seventeen noise scales in Fig.~\ref{fig:sweep}. Their difference is the override effect. Intervals are the corresponding binomial marginal bounds from the retrospective $27\times10$ display family. These are the same samples as in the dense sweep, not an independent experiment.}
\label{fig:repairdecomposition}\end{figure}

For a sampled record and logical label, define
\begin{equation}
X_j=\mathbf 1\{D_0(S_j)\ne L_j\}-\mathbf 1\{D_1(S_j)\ne L_j\}.
\end{equation}
Only discordant outcomes contribute: $X_j=1$ for a repair, $-1$ for an
introduced failure, and zero otherwise. The decoders agree outside
$\mathcal A$, so all nonzero contributions arise from the selected complete
records. Figure~\ref{fig:repairdecomposition} resolves the two probabilities
whose difference is plotted in Fig.~\ref{fig:sweep}. It uses the same samples
and does not add independent evidence.

The retained records also resolve each target separately and give the
absolute failure rates of both decoders. Those additional views are omitted here because they do not change the
reversal or narrow its confidence bracket.

\section{Paired sampling and simultaneous intervals}
\label{app:statistics}
For the original sweep and intervention conditions, we retain four binomial marginals: ordinary decoder
failure, modified decoder failure, repair of an ordinary failure, and
introduction of a failure by the modification. If $r$ and $h$ are the repair
and introduced-failure probabilities, then $\Delta=r-h$. Given marginal
intervals, the effect interval is $[r_L-h_U,r_U-h_L]$.

The initial noise-sweep family allocated failure probability
$0.05/(7\cdot4)$ per marginal: six noise scales and one targeted-suppression
condition, each with twenty million samples. The independent initial
attenuation/control family used $0.05/(6\cdot4)$. Their original results
remain separate from the later display reanalysis.

Figures~\ref{fig:sweep}, \ref{fig:matched-2}, \ref{fig:repairdecomposition},
\ref{fig:matched}, and \ref{fig:matched-3} use a common retrospective family
of 27 conditions: seventeen noise scales, nine attenuation fractions, and
one matched suppression. Ten binomial marginals per condition comprise
the four decoder/discordance counts and the two logical-label counts for
each of the three selected syndromes. We allocate $0.05/(27\cdot10)$
per marginal and recompute Clopper--Pearson bounds \cite{clopper}.
The display contains 540 million shots: 240 million retained from the
initial experiments and 300 million from the eleven additional noise
scales and four additional attenuation fractions. The extension grid was
chosen after the initial results and fixed before its samples were drawn.
The combined display is therefore a retrospective reanalysis, not a
preregistered confirmation experiment.

For effects with bounds $[L_i,U_i]$ and $[L_j,U_j]$, the contrast is bounded
by $[L_i-U_j,U_i-L_j]$. No independence assumption is needed for this interval
subtraction. The broader family widens the intervals without changing the
retained point estimates. The control baseline remains independent of the
noise-sweep baseline; samples are not pooled between them.

The joint-likelihood, mixed-bias, and temporal-boundary experiments retain
their separate confidence families, as stated in their captions. These
families are not combined into a study-wide 95\% claim. Sampling uncertainty
does not include error from physical-model misspecification.

\section{Target and control locations}
\label{app:intervention}
\label{app:controls}
The original instruction is matched for every target--control pair, fixing
channel type, probability, tick, extraction round, and surgery phase.
Within each instruction, controls minimize total Manhattan separation;
lexicographic qubit-index tie-breaking makes selection deterministic.
Same-qubit reset/readout pairings are preserved. Table~\ref{tab:controls}
lists the locations in the original circuit coordinates.
\begin{table}[b]
\caption{Target and matched-control qubits. Each row identifies both the
post-reset channel and the later readout-flip channel at the stated ticks.}
\label{tab:controls}
\begin{ruledtabular}
\begin{tabular}{ccc}
Reset/readout ticks & Target qubits & Control qubits\\
\hline
20/25 & 27, 39 & 1, 13\\
32/37 & 1, 13 & 0, 2
\end{tabular}
\end{ruledtabular}
\end{table}
Every spatial separation is two coordinate units. Controls are non-target
locations, not disjoint qubit identities: the same qubit may appear in
different roles at different times. Independent enumeration confirms that
no control location participates in an identified label-one reset/readout
pair for a selected syndrome. This does not match their full detector
supports, graph degrees, or boundary roles. The audit verifies equality of
the ideal circuits and of all unmodified physical noise channels.

\subsection{Attenuation and derived contrasts}
\begin{figure}[t]
\centering
\includegraphics[width=\columnwidth]{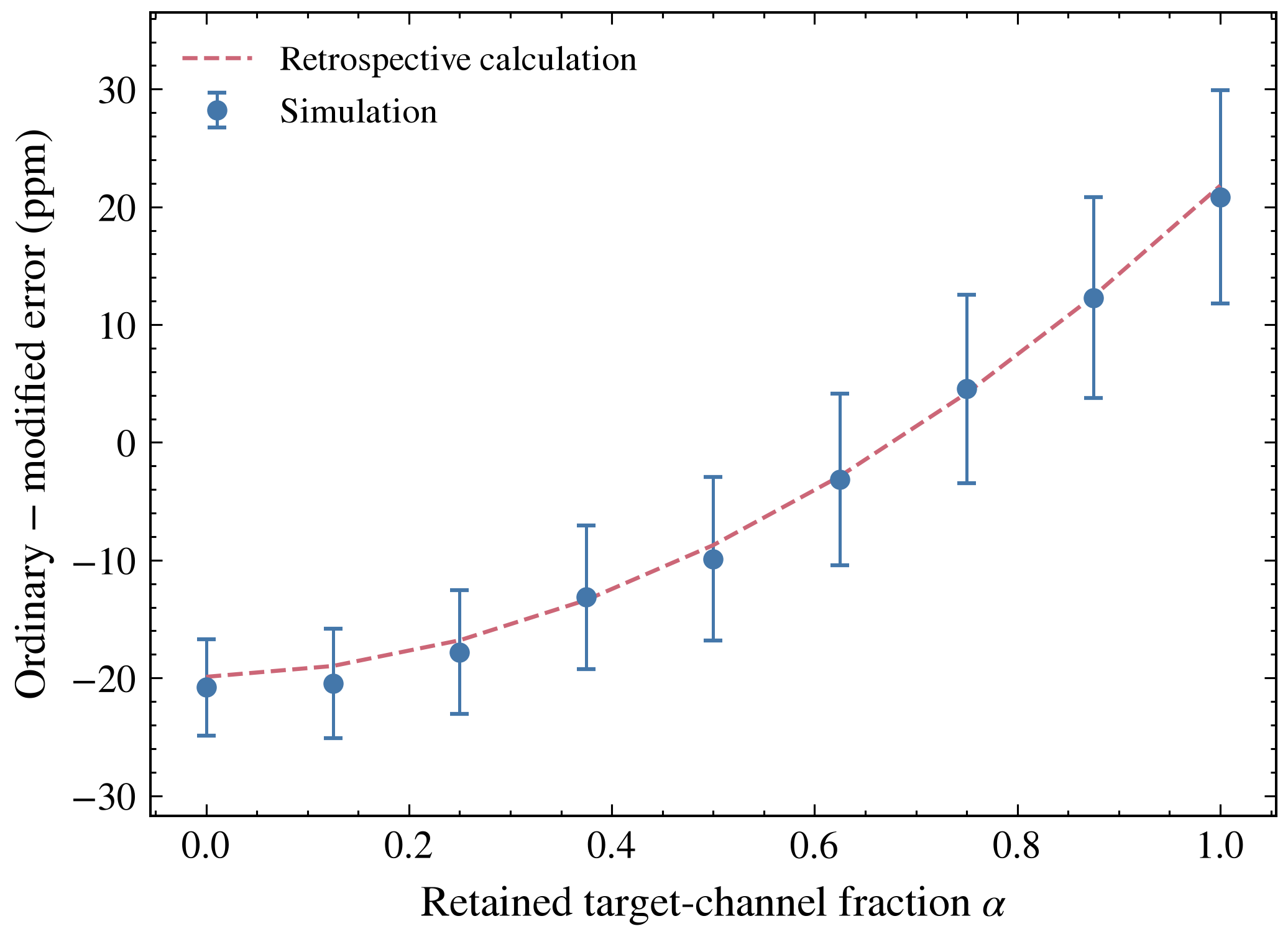}
\caption{Targeted attenuation at $\gamma=0.50$, with nine retained fractions and twenty million shots per condition. Only the eight selected channel probabilities are multiplied by $\alpha$. Five fractions reuse the initial experiment; four were added later. Intervals use the retrospective $27\times10$ display family. The dashed curve is an unfitted, retrospective calculation of $\Delta(\alpha)$ with all local fault orders and at most two background faults. It was evaluated after the observations and supplies a model check, not prospective confirmation.}
\label{fig:matched}\end{figure}

\begin{figure}[t]
\centering
\includegraphics[width=\columnwidth]{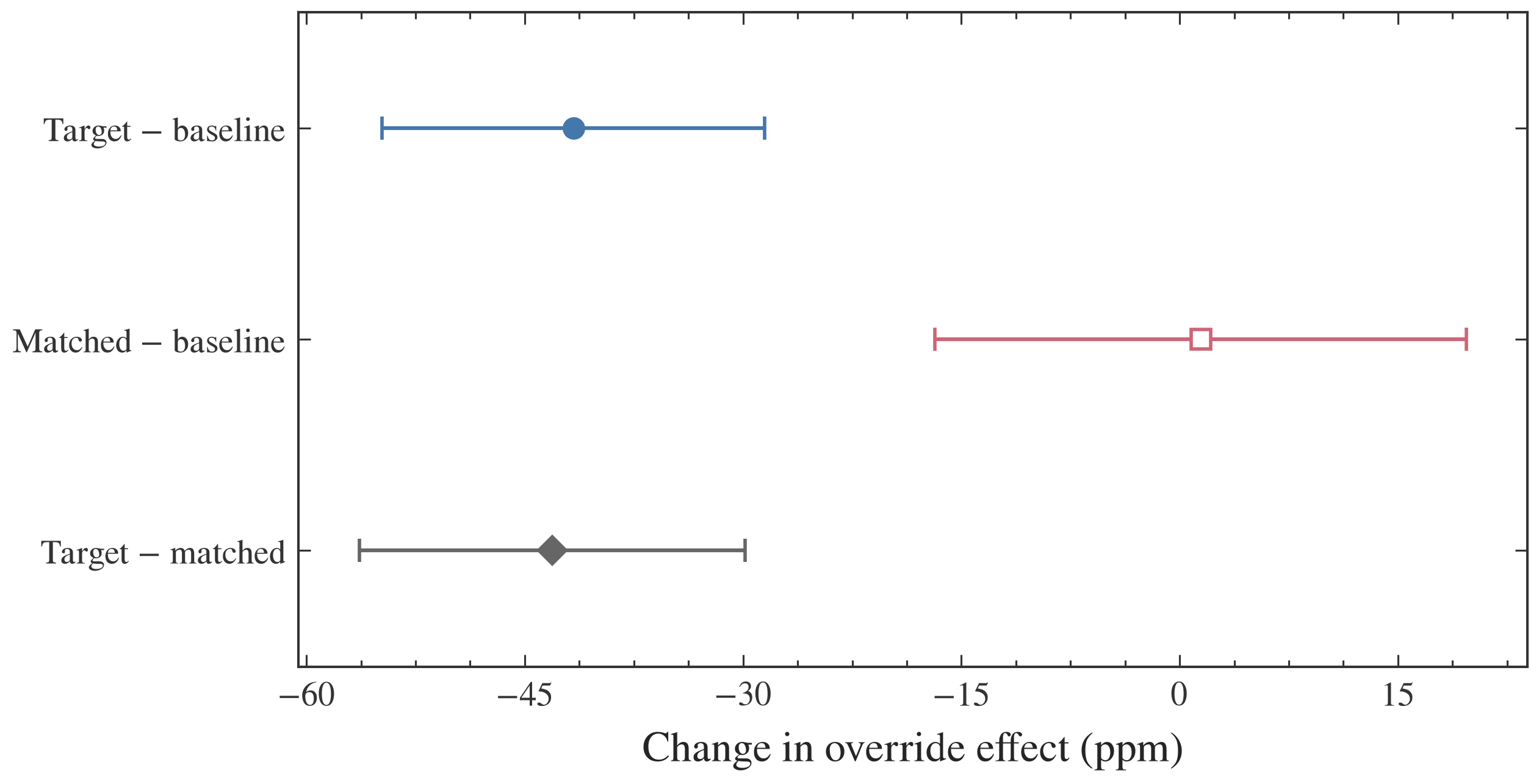}
\caption{Changes in override effect for targeted minus baseline, matched minus baseline, and targeted minus matched suppression. The bounds are differences of interval endpoints from the retrospective $27\times10$ display family. The unresolved matched-minus-baseline contrast is not an equivalence result.}
\label{fig:matched-3}\end{figure}

Figure~\ref{fig:matched} reports the attenuation series, including the two
endpoints used in the main intervention comparison. Figure~\ref{fig:matched-3}
shows contrasts obtained by subtracting the corresponding simultaneous
interval endpoints. The attenuation plot combines the five original
fractions with four later fractions, using the retrospective display family.
The contrasts reuse the original baseline and two suppression conditions;
they introduce no additional samples or independence assumption.

\section{Circuit and numerical validation}
\label{app:validation}
The FTPrimitiveBench revision is
\texttt{b75fbd1c514bc4fb7fcca75618a191e6b58fa7ed}.
The full circuit is simulated with Stim \cite{stim}; matching uses PyMatching
\cite{pymatching}. Independent reference-entangled ideal-operation checks
sample 128 trajectories at each of $d=3$ and $d=5$, including both parity
branches and negative controls. These are sampled ideal-operation checks,
not exhaustive validation of the noisy logical instrument.

Backward Pauli propagation checks the physical witness signatures without
using the production detector-error explanations. Separate characteristic-
function products and inverse Walsh transforms reconstruct the integrated
local laws. Confidence endpoints are recomputed through a different
beta-inverse interface. Channel multisets, paired count identities, and
target-label totals are checked, with source hashes and sampling seeds retained.
These checks share the underlying Stim simulator; they are not independent
full-simulator implementations or rigorous enclosures of all floating-point
operations.

Replaying 27 verification procedures and 12.6 million original seeded
samples checks the retained results. Fixed-seed agreement establishes
reproducibility, not fresh statistical confirmation. The retained records
include the full-syndrome likelihood calculation, the direct parity-error
comparison, and the targeted and matched interventions, together with
sampling seeds and source hashes.

\subsection{Verification of single-fault decisions}
An earlier implementation of the mixed-bias and correlated-matching
single-fault checks used numeric selection masks as row indices. This
selected the wrong rows. The corrected procedure uses Boolean masks and
checks membership against complete integer-encoded detector signatures.
All reported single-fault counts use the corrected procedure, which was
checked with both numeric and Boolean signature arrays across all twenty-one
noise conditions. The correction changes the affected single-fault counts.
It does not change Monte Carlo failure counts or confidence intervals:
the sampler already supplied Boolean arrays. The temporal-boundary
experiment used the corrected procedure from the outset.

The mixed-bias predictions, correlated-matching comparisons, and
preparation--recovery extension each retain their prediction records and
verification results. Their confirmation families allocate
$0.05/(21\cdot14)$, $0.05/(21\cdot16)$, and $0.05/(3\cdot4)$ per marginal,
respectively. These are separate confidence statements. Predictions were
saved locally before sampling, not publicly preregistered.

\section{Forecasts and scope of transfer}
\label{app:forecasts}
For the original depolarizing sweep, before direct confirmation, an all-order conditional-likelihood calculation
with two million background samples per condition supplied frozen point
forecasts. Their predicted crossing lies between $0.20$ and $0.25$, and the
forecasts agree with the later direct intervals. However, every conservative
simultaneous forecast interval contains zero. They do not certify that
crossover location.

The leading-history balance and the full quadratic expansion give different
numbers. For $s=D_{63}D_{77}$, the two single-gate histories contribute
$A_0=1.3333333\times10^{-4}$, while the four dominant reset/readout pairs
contribute $5.76\times10^{-4}$ to the quadratic coefficient of label one.
Equating only these contributions gives
$\gamma=A_0/(5.76\times10^{-4})=0.23148$.
The full expansion $q(s,\ell)=A_\ell\gamma+B_\ell\gamma^2+O(\gamma^3)$
also includes the probability that every other location is fault-free:
\begin{align}
 B_0 &= -6.5391111\times10^{-4}+1.7005432\times10^{-5}
       \nonumber\\
     &= -6.3690568\times10^{-4},\\
 B_1 &= 6.0705235\times10^{-4}.
\end{align}
The negative term in $B_0$ is the single-fault survival correction; the
positive term counts distinct-location pairs carrying label zero.
Consequently $A_0/(B_1-B_0)=0.10718$, with roots $0.10718$, $0.10896$,
and $0.10718$ for the three records. Independent ordered-pair enumeration,
subtracting mutually exclusive same-location alternatives and dividing by
two, reproduces these coefficients. There is no factor-of-two discrepancy
in the pair count. The two estimates retain different terms, and neither
has a finite-noise remainder guarantee. The agreement of $0.23148$ with the
all-order crossing window is therefore a useful heuristic, not a validation
of the quadratic truncation.

\subsection{Reset/readout ratios at fixed gate noise}
An existing nine-condition study holds $p_{2q}=5\times10^{-4}$,
$p_{1q}=5\times10^{-5}$, and both idle rates at $5\times10^{-4}/6$.
It varies $u=p_{\rm reset}+p_{\rm meas}$ and
$f=p_{\rm reset}/u$. Table~\ref{tab:rate-ratios} expresses this grid as
ratios to $p_{2q}$. Each condition uses five million background samples
with all local fault orders integrated. The simultaneous family contains
$9\times32768$ marginal bounds. All three selected syndromes favor label
zero at $u=0.004$, are unresolved at $0.008$, and favor label one at
$0.012$, for every tested split. Thus changing gate noise is not necessary
for the reversal. This grid varies reset and measurement rates together;
it is not a measurement-only sweep at fixed reset noise.
\begin{table}[t]
\caption{Controlled rate grid. Each entry lists
$(p_{\rm reset}/p_{2q},p_{\rm meas}/p_{2q})$.}
\label{tab:rate-ratios}\centering
\begin{tabular}{lccc}\toprule
$u$ & $f=0.25$ & $f=0.50$ & $f=0.75$\\\midrule
0.004 & $(2,6)$ & $(4,4)$ & $(6,2)$\\
0.008 & $(4,12)$ & $(8,8)$ & $(12,4)$\\
0.012 & $(6,18)$ & $(12,12)$ & $(18,6)$\\\bottomrule
\end{tabular}
\end{table}
Independent interval recomputation checks the retained rate-grid results.
The eighteen contrasts between
unequal and balanced splits at fixed $u$ all include zero; this does not
establish equivalence. Paired reset/readout channels with the same complete
signature combine with probability
$p_{\rm reset}+p_{\rm meas}-2p_{\rm reset}p_{\rm meas}$, but unmatched
preparation and terminal-readout channels prevent a general invariance
under changing the split.

\subsection{Additional circuit tests}
\label{app:transfer}
A structural transfer test increases the merged-round count from three to
four. A candidate is a complete two-detector record with at least one
single-location explanation, all such explanations carrying label zero,
and at least one distinct-location reset/readout pair carrying label one.
Let $A_0(s)$ be the sum of its single-fault coefficients and
$B_{\mathrm{rr},1}(s)$ the sum of coefficient products for those unordered
reset/readout pairs. Coefficients use the reference noise model at
$\gamma=1$. Candidates are ranked by decreasing
$B_{\mathrm{rr},1}(s)/A_0(s)$, with ties broken lexicographically by the
ordered detector indices. The first three candidates are selected.
This is a structural selection rule, not an assertion of likelihood
optimality.
It finds nine candidates at three rounds and none at four rounds. Independent
physical propagation confirms the negative result. This excludes the
specified construction in the tested four-round circuit, not every possible
finite-noise trade-off. This extension has sampled ideal-detector checks but
no complete independent reference-entangled operation audit.

A separately archived $d=5$ selected two-fault witness has a positive
joint-mass-difference interval $[1.1882,4.4589]\times10^{-11}$ from a fixed
pivot estimator. Its small full-syndrome probability, different selected
event, and limited protection audit prevent interpreting it as a
distance-scaling curve or a complete two-fault-correction certificate.
A high-precision diagnostic bounds one latent-probability representation
discrepancy contribution by $1.313\times10^{-12}$, without enclosing all
numerical operations. This is supporting evidence only. The auxiliary witness concerns a
different competition from the single-fault-versus-reset/readout-pair
construction used in the structural selection test.

The study originated in reset placement. An exhaustive temporal-schedule
pilot did not resolve a mixed-reset advantage at its tested physical noise
point. That negative benchmark and the preceding toy models are retained
in the study records. They supply no reset-saving claim for the present work.

\subsection{Mixed-noise prediction model}
For two-qubit faults, probabilities $0.2,0.2,0.6$ select support on the first
qubit, second qubit, or both. Conditional axes on two occupied qubits are
independent draws from the specified $X,Y,Z$ weights. Total gate and idle
fault probabilities are held fixed, while reset/readout channels retain
their basis-specific form. This construction is isotropic at equal weights.
Direct circuit sampling preserves mutually exclusive outcomes at each
physical location; graph construction uses Stim's disjoint-error
approximation for the mixed channels.

The deterministic prediction integrates every fault order from channels
supported in the selected local region. Remaining channels contribute through
exact zero-, one-, and two-fault terms with zero outside syndrome. Pair sums
exclude two alternatives from the same location, and truncated mass is not
renormalized. Independent Fourier-product reconstruction verifies the local
law, and direct enumeration verifies distinct-location background pairs.
These are separate calculations of the same model, not independent physical
simulators or rigorous floating-point enclosures.

For the extended preparation/recovery circuit, the selected records are
$D_{79}D_{93}$, $D_{79}D_{96}$, and $D_{82}D_{96}$, with all other bits zero.
Their 14-detector support translates that of the original witnesses. The
unconditioned omitted-background mass reaches approximately $0.516$ at
$\gamma=0.50$. Its bound is too broad to certify the useful forecast sign;
the reported agreement is empirical.

\subsection{Correlated-matching replay}
\label{app:correlated}
Two-pass correlated matching \cite{fowlercorrelated} is enabled in graph
construction and decoding. Unlike comparisons that exploit correlations
between blocks in transversal circuits \cite{cain}, this test holds the
available record fixed and changes only the decoder and the selected labels.
The same 420 million mixed-noise records are replayed. The ordinary-decoder
failure counts and target counts reproduce the original run exactly.

Adding the frozen decisions to correlated matching gives six resolved
benefits and twelve exact identities in the eighteen mixed-bias conditions.
Including the three isotropic controls gives seven benefits, thirteen
identities, and one unresolved effect. At the isotropic $\gamma=0.50$ control,
correlated matching makes 111281 errors and its modified version 110902 in
twenty million shots. The reduction is $18.95$ ppm, with simultaneous bounds
$[9.70,28.20]$ ppm. The baseline target labels agree for ordinary and correlated
matching in all twenty-one conditions; identical incremental effects on
shared samples follow algebraically. Their global decisions differ, and
neither decoder is uniformly better. This heuristic comparison does not
exhaust likelihood-aware baselines or supply an independent replication.

\subsection{Fully enlarged circuit and symmetry checks}
\label{app:fully-enlarged}
The fixed selection rule finds no eligible two-detector single-fault versus
reset/readout-pair collision in the circuit with distance-five patches and
five rounds in each stage. Independent propagation agrees with Stim
explanations for all 50378 physical alternatives. The extended-boundary and
fully enlarged circuits each pass 128 reference-entangled ideal trajectories
and 1024 noiseless detector samples. The checks use the existing geometric
validation logic; they do not prove an exact noisy fault distance.

Full Hadamard conjugation transforms preparations, measurements, gates,
Pauli channels, and logical frames together. The mapped physical signatures
and ideal operation pass the corresponding checks. Replaying 2.1 million
dual-circuit records gives no sampled disagreement between transported and
rebuilt decoders for either tested matching mode. This finite-sample symmetry
check supplies no independent mechanism transfer and is not a comparison
against an unrelated native $ZZ$ circuit.

\end{document}